\documentclass{article}
\usepackage{kmn}
\usepackage{epsfig,subfigure}
\title[The Galactic Disk Distribution of Planetary Nebulae With Warm
Dust Emission Features: I]{The Galactic Disk Distribution of Planetary
Nebulae With Warm Dust Emission Features: I}

\author[S.~Casassus et al.]
  {S.~Casassus$^{1,2}$, P.\,F.~Roche$^1$, D.\,K.~Aitken$^3$ \& C.\,H.~Smith$^4$ \\
	$^1$ Astrophysics, Physics Department, Oxford University,
  Keble Road, Oxford OX1 3RH\\
 	$^2$ Departamento de Astronom\'{\i}a, Universidad de Chile,
  Casilla 36-D, Santiago, Chile.\\
	$^3$ Department of Physical Sciences, University of Hertfordshire, Hatfield, Herts AL10 9AB\\
	$^4$ School of Physics, University College, UNSW, Canberra,
  ACT 2600, Australia.}
\date{Accepted ... Received ...}
\pagerange{\pageref{firstpage}--\pageref{lastpage}}
\pubyear{2000}
\def\gs{\mathrel{\raise1.16pt\hbox{$>$}\kern-7.0pt
\lower3.06pt\hbox{{$\scriptstyle \sim$}}}}
\def\ls{\mathrel{\raise1.16pt\hbox{$<$}\kern-7.0pt
\lower3.06pt\hbox{{$\scriptstyle \sim$}}}}
\begin{document}
\label{firstpage}
\maketitle


\begin{abstract}
We investigate the galactic disk distribution of a sample of planetary
nebulae characterised in terms of their mid-infrared spectral
features. The total number of galactic disk PNe with 8--13$\mu$m
spectra is brought up to 74 with the inclusion of 24 new objects,
whose spectra we present for the first time.  54 PNe have clearly
identified warm dust emission features, and form a sample which we use
to construct the distribution of the C/O chemical balance in galactic
disk PNe. The dust emission features complement the information on the
progenitor masses brought by the gas-phase N/O ratios: PNe with
unidentified infrared emission bands have the highest N/O ratios,
while PNe with the silicate signature have either very high N
enrichment or close to none, and SiC emission features coincide with a
range of moderate N-enrichments. We find a trend for a decreasing
proportion of O-rich PNe towards the third and fourth galactic
quadrants. Two independent distance scales confirmed that the
proportion of O-rich PNe decreases from 30$\pm$9\% inside the solar
circle, to 14$\pm$7\% outside.  PNe with warm dust are also the
youngest. PNe with no warm dust are uniformly distributed in C/O and
N/O ratios, and do not appear to be confined to C/O$\sim$1.  They also
have higher 6\,cm fluxes, as expected from more evolved PNe. We show
that the {\em IRAS} fluxes are a good representation of the bolometric
flux for warm-dust PNe. The requirement $F(12\mu$m)\,$>$\,0.5\,Jy
should probe a good portion of the galactic disk, and the dominant
selection effects are rooted in the PN catalogues.
\end{abstract}

\section{Introduction}


Spectroscopy at 10\,$\mu$m has brought significant information on the
chemical composition of planetary nebulae (PNe). The dust signatures
reflect the C/O chemical balance at the tip of the asymptotic giant
branch (AGB). In this article we use the 8--13$\mu$m dust signatures
as a systematic tool to investigate the distribution of the C/O
abundance ratio in PNe. We classify the 8--13$\mu$m spectral
signatures in a sample consisting of compact and IR-bright PNe in the
Strasbourg-ESO catalogue (Acker et al. 1992). The sample excludes PNe
traditionally associated with the galactic bulge, which will make the
object of a forthcoming article.

The family of emission bands usually referred to as the `unidentified
infrared bands' (UIR bands), with principal members at 3.3$\mu$m,
6.2$\mu$m, 7.7$\mu$m, 8.6$\mu$m and 11.3$\mu$m, were first observed in
the mid-IR towards \mbox{ NGC\,7027} by Gillett et al. (1973).  Cohen
et al. (1986, 1989) found good correlations between the strengths of
all pairs of bands towards a sample of PNe, reflection nebulae and HII
regions, showing that they correspond to a generic spectrum. The
strength of the 7.7$\mu$m feature relative to the total {\em IRAS} flux
correlates strongly with the gas phase C/O ratio in a sample of 6 PNe
(Cohen et al. 1986), and Duley \& Williams (1981) identified the
principal wavelengths of the UIR bands with transitions in the
chemical functional groups of aromatic molecules. Polycyclic aromatic
hydrocarbons are the commonly accepted carriers for the UIR bands
(L\'{e}ger \& Puget 1984). Thus, although their exact carriers still
remain to be determined, the UIR bands are indicative of a carbon rich
environment.

The 8--13$\mu$m spectra of PNe can also show significant continuum
emission attributed to `warm' dust at $\sim$200\,K.  As first shown by
Aitken et al. (1979), a smooth emission feature with a peak at about
9.7$\mu$m is observed in the PNe SwSt\,1, M\,1-26 and Hb\,12. The peak
of emission coincides with the Si-O stretch in silicates (Day 1979,
1981), and this feature is similar to amorphous condensates of
silicate materials (Day and Donn 1978). It is typical of the Trapezium
region in Orion and of the circumstellar shells of some oxygen-rich
stars (Forrest et al. 1975), and is also seen in absorption towards the
BN infrared point source in Orion (Gillett et al. 1975). Willner at
al. (1979) detected a smooth emission feature with a rather flat
profile, from 10.5$\mu$m to about 12.7$\mu$m, towards the PNe IC\,418
and NGC\,6572. It is commonly observed towards C stars as excess
emission over a black body spectrum, and is attributed to lattice
vibrations in silicon carbide (Forrest et al. 1975). Andersen et
al. (1999) published the mid-IR transmission spectrum of meteoritic
SiC grains; they obtained a good match to the C star feature when
extracting grains with a small size distribution (i.e. $<5\mu$m).

There are thus significant compositional differences in the dust
content of PNe, which can be classified into dust emission types, and
according to whether they contain O-rich or C-rich grain materials
(Aitken et al. 1979). How do the 10$\mu$m dust emission features
compare with the gas phase abundances? What is the proportion of PNe
that show each type of dust emission, and does their distribution show
large scale variations across the galactic disk?  It has been shown by
Thronson et al. (1987) and Jura et al. (1989) that the proportion of C
stars relative to M giants increases outside the solar circle, and it
is interesting to investigate whether PNe follow a similar trend. In
contrast to the C and M stars which span a range of locations in the
giant branch, PN compositions reflect the surface abundances at the
end of the AGB, with a well defined evolutionary status.

This article is the first of a series devoted to the statistical
analysis of the grain composition in galactic disk PNe.  In
Section~\ref{sec:dust_obs} we present a sample of 74 PNe with
8-13$\mu$m spectra, which includes 24 previously unpublished spectra
obtained with CGS3 on UKIRT or the UCL spectrometer on UKIRT or the
AAT.  Section~\ref{sec:dust_obs} also contains a brief description of
the method used to classify the 10$\mu$m continua (following Aitken et
al. 1979).  We will then compare in Section \ref{sec:gasphase} the
dust content of PNe with their gas phase C/O and N/O abundance ratios,
to show that the dust emission types represent an alternative for
determining the C/O chemical balance and bring complementary
information on the PN progenitors.  The sky distribution of PN dust
types is presented in Section \ref{sec:sky}. After adopting a
statistical distance scale in Section \ref{sec:PNdists}, we will test
in Section \ref{sec:statproof} the size of the PN sample in this work
for stratification in Peimbert (1978) types, and discuss its
homogeneity, giving some support for its statistical significance. The
galactic disk distribution of the PN dust composition is presented in
Section \ref{sec:pnedistrib}.  Section \ref{sec:conclusion_chap3}
summarises our conclusions.

\section{8--13${\mu}$\lowercase{m} spectroscopy of compact and infrared-bright PNe}\label{sec:dust_obs}

\begin{figure*}
\resizebox{17cm}{!}{\epsfig{file=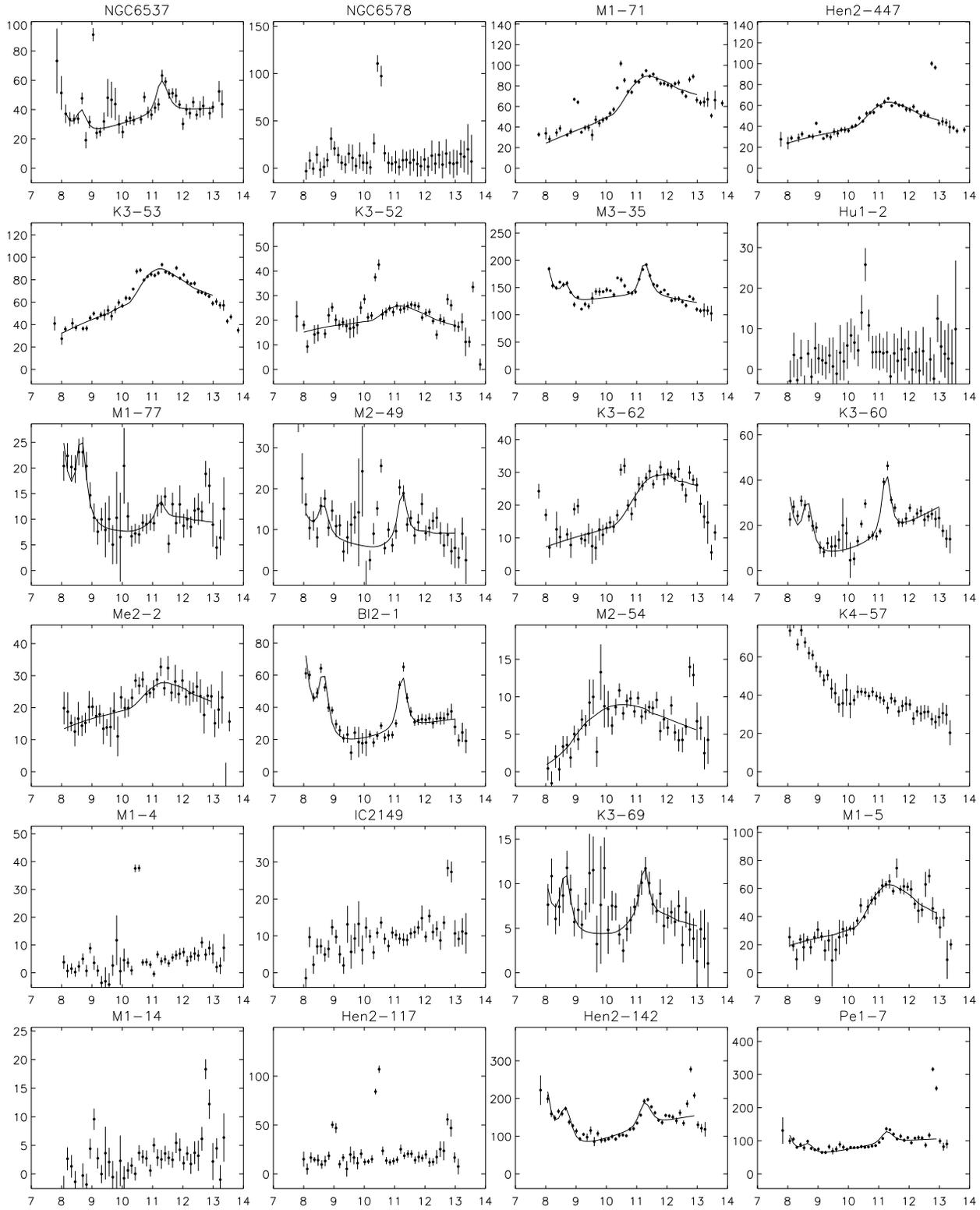}}
\caption{10$\mu$m spectra for the PNe listed in Table
\ref{table:obslog}. In abscissae is the wavelength range in $\mu$m,
and in ordinates the flux density in
10$^{-19}$~W\,cm$^{-2}$\,$\mu$m$^{-1}$. A fit to the continuum
emission is shown for the spectra with identified dust features, with
parameters listed in Table \ref{table:dusttypes}. The data points
around bright emission lines ([A\,{\sc iii}], [S\,{\sc iv}] and
[Ne\,{\sc ii}] at 8.99, 10.52 and 12.81 $\mu$m respectively) are
excluded from the fits.}\label{fig:dustspecs}
\end{figure*}

The criteria for selection of the PN sample were that they be
compact, less than 10$''$ in diameter so that most of the flux is
contained in the spectrograph beam, and infrared bright, with {\em
IRAS} 12$\mu$m flux in excess of 0.5~Jy. These criteria select the
best candidates for the detection of the dust emission features. A
sample of 10 PNe was observed with CGS3 on UKIRT on 1996 September 27
and 28, with the intention of increasing the number of objects
measured beyond the solar circle. An observing log is presented in
Table~\ref{table:obslog} where details of the previously unpublished
spectra from the AAT and UKIRT are listed.  Both CGS3 and the UCLS
were used in their low-resolution modes, producing oversampled spectra
which are calibrated with respect to standard stars. Fluxes are
accurate to 20\%.  The fluxes of emission lines present in some
objects are listed in table \ref{table:fluxes}. Although they provide
information on the ionized gaseous component, we will not use the
emission lines for the purpose of this work. The resulting spectra are
shown in Figure \ref{fig:dustspecs}, together with fits to the
8-13$\mu$m continua based on the grain emissivities, which form the
basis of the dust type classification.


\begin{table}
\caption{Log of observations.}\label{table:obslog}
\begin{center}
\small
\begin{tabular}{lrlcc}    \hline
 	         &  $l$ &  $b$       &  instrument  &       Date    \\
                 &\multicolumn{2}{c}{[degrees]}&       & 	\\
        NGC6537  & 10.01 &  0.70 & UKIRT+CGS3  &  May 94        \\
        NGC6578  & 10.80 & -1.80 & UKIRT+UCLS  &  Oct 87 	\\
          M1-71  & 55.50 & -0.50 & UKIRT+UCLS  &  Jul 90	\\
       Hen2-447  & 57.90 & -1.50 & UKIRT+UCLS  &  Jul 90        \\
          K3-53  & 64.90 & -2.10 & UKIRT+UCLS  &  Jul 90	\\
          K3-52  & 67.90 & -0.20 & UKIRT+UCLS  &  Jul 90	\\
          M3-35  & 71.60 & -2.30 & UKIRT+UCLS  &  Jul 86	\\
          Hu1-2  & 86.50 & -8.80 & UKIRT+UCLS  &  Oct 87	\\
          M1-77  & 89.30 & -2.20 & UKIRT+CGS3  &  Sep 96	\\
          M2-49  & 95.10 & -2.00 & UKIRT+CGS3  &  Sep 96  	\\
          K3-62  & 95.20 & +0.70 & UKIRT+UCLS  &  Jul 90	\\
          K3-60  & 98.20 & +4.90 & UKIRT+CGS3  &  Sep 96	\\
          Me2-2  & 100.00& -8.70 & UKIRT+UCLS  &  Oct 87	\\
          Bl2-1  & 104.10& +1.00 & UKIRT+CGS3  &  Sep 96	\\
          M2-54  & 104.80& -6.70 & UKIRT+CGS3  &  Sep 96	\\
          K4-57  & 107.40& -0.60 & UKIRT+CGS3  &  Sep 96	\\
           M1-4  & 147.40& -2.30 & UKIRT+CGS3  &  Sep 96	\\
         IC2149  & 166.10&+10.40 & UKIRT+CGS3  &  Sep 96	\\
          K3-69  & 170.70& +4.60 & UKIRT+CGS3  &  Sep 96	\\
           M1-5  & 184.00& -2.10 & UKIRT+UCLS  &  Oct 87	\\
          M1-14  & 234.90& -1.40 & UKIRT+CGS3  &  Sep 96	\\
       Hen2-117  & 320.90& +2.20 & AAT+UCLS    &  Apr 87        \\
       Hen2-142  & 327.10& -2.20 & AAT+UCLS    &  Apr 86 	\\
          Pe1-7  & 337.40& +1.60 & AAT+UCLS    &  Apr 86 	\\  \hline
\end{tabular}
\end{center}
\end{table}
\begin{table}
\caption{Emission line fluxes, in 10$^{-15}$W m$^{-2}$ .}\label{table:fluxes}
\begin{center}

\begin{tabular}{lccc}   \hline
           & [Ar\,{\sc iii}] & [S\,{\sc iv}] & [Ne\,{\sc ii}] \\
           &  8.99\,$\mu$m   & 10.52\,$\mu$m & 12.81\,$\mu$m \\
   NGC6578 & 2.10    &   17.0     &    ---   \\
     M1-71 & 3.57    &   5.52     &  2.27    \\
  Hen2-447 & 1.04    &    ---     &  5.98    \\
     K3-53 &   ---   &   2.56     &    ---   \\
     K3-52 & 0.51    &   2.45     &  1.00    \\
     M3-35 &   ---   &   4.77     &  1.99   \\
     Hu1-2 &   ---   &   1.93     &    ---   \\
     M1-77 &   ---   &     ---    &  0.98    \\
     M2-49 &   ---   &   2.29     &    ---   \\
     K3-62 & 1.16    &   2.36     &  0.58    \\
     K3-60 &   ---   &   2.27     &    ---   \\
     Me2-2 &   ---   &   0.63     &    ---   \\
     M2-54 &   ---   &     ---    &  0.90    \\
      M1-4 &   ---   &   5.52     &    ---   \\
    IC2149 &   ---   &     ---    &  2.02    \\
     M1-14 & 0.89    &     ---    &  1.33    \\
  Hen2-117 & 4.03    &   13.9     &  3.92    \\
  Hen2-142 &   ---   &     ---    &  11.8     \\
     Pe1-7 &   ---   &     ---    &  22.5   \\   \hline
\end{tabular}
\end{center}
\end{table}



We classified the type of dust emission features according to the
procedure described in Aitken et al. (1979) and Aitken and Roche
(1982).  The emissivity functions $\epsilon_i$ for the three types of
grains are taken from the spectra of astrophysical sources, with
\mbox{$F_{\lambda}=\epsilon_i\,B(\lambda,T)$}, where $F_{\lambda}$ is
the observed flux density and $B(\lambda,T)$ is a Planck function at
temperature $T$ ($\epsilon_i$ is then fixed by
$\epsilon_i(10\mu$m)=1). Additionally a smooth continuum with
emissivity $f(\lambda)\propto \lambda^{-1.8}$, which is taken to
represent graphite grains (or amorphous carbon grains), was included
in the fitting procedure.  The relative contribution of emissivity
functions and the temperatures of the grains are fit to the 8-13$\mu$m
continua following a $\chi^{2}$ minimisation:
\begin{eqnarray}
F_{\lambda}={\sum}_{i} a_{i} \epsilon_i({\lambda})
B({\lambda},T_{i})/B(10{\mu}m,T_i), \\ \label{eq:Fl}
\chi^2=\sum_\lambda (F_{\lambda}^\mathrm{obs}-F_{\lambda})^2/{\sigma_{\lambda}}^2,
\end{eqnarray}
where the sum extends to the number of dust components used in the
fit, and $\epsilon_{\lambda}$ stands for the error in each point of
the observed spectrum $F_{\lambda}^\mathrm{obs}$.  The Planck
functions are arbitrarily normalised at 10$\mu$m. Best fit values are
thus obtained for the measure of contribution of each emission type
$a_{1-4}$, and their black body temperatures $T_{1-4}$.  The relative
contribution of each dust type is
$a^{\prime}_{i}=a_{i}/\sum_{j}a_{j}$, summing over all the dust types
required in the fit. The coefficient a$^{\prime}$ is thus a
representation of the fractional emission at $\lambda$=10$\mu$m.  The
fitting procedure assumes that all emitting materials are optically
thin and that any absorption occurs in a cold foreground layer; in the
case of absorption a factor $e^{-\tau(\lambda)}$ is included in
Eq. \ref{eq:Fl} ($\tau(\lambda)$ is the wavelength dependent opacity,
with a profile given by the emissivity curve, keeping $\tau$(10$\mu$m)
as a free parameter).  The available data provide insufficient
constraints to warrant a more detailed radiative transfer treatment.

As discussed in Aitken and Roche (1982), it is the silicate grain
emission feature that really separates PNe into different groups. A
portion of PNe with silicate emission also show the UIR bands, and
sometimes require SiC in small amounts. Graphite emission seems to be
unrelated to the other types of grains. The PN dust signatures can be
placed into four groups based on the dominant dust species at 10
$\mu$m as shown in Table~\ref{table:class}.


\begin{table}
\caption{Classification of PN dust types.}\label{table:class}
\begin{center}
\begin{tabular}{ccl}
O & (silicates) & Silicate grain emission.\\
C & (UIR) & UIR bands but not Silicates.\\
c & (SiC) & SiC grains, but neither Silicates \\
  &       & nor the UIR bands.\\
+ & (weak) & weak or unidentifiable continuum.
\end{tabular}
\end{center}
\end{table}

A list of all the identified dust emission features found in this
sample of PNe can be found in Table \ref{table:dusttypes}. The S/N
ratio of many spectra are rather low, and higher quality data may
confirm the need for mixture of grain types. However, in this article
we are concerned with the dominant material. The column under
`comments' gives more information on the best fit parameters, in the
form
\begin{equation}\label{eq:notation}
(\mathrm{grain type}):a^{\prime},T,\tau   ,
\end{equation}
where the optical depth field $\tau$ is listed for only when
absorption is required (which is the case for M\,2-9, M\,2-56 and 19w32).

As would be expected from blackbody radiation between 8-13$\mu$m, a typical
dust temperature is $\gs$\,200K, and the 8-13$\mu$m dust emission can
be referred to as `warm dust emission features' in comparison with
colder dust $\ls$200K which makes the bulk of the FIR emission
(e.g. Kwok et al. 1986).

\begin{table*} 
 \vbox to220mm{\vfil Landscape Table to go here
 \caption{}  \label{table:dusttypes}
 \vfil}
\end{table*}

The objects are listed in the Strasbourg-ESO catalogue (Acker et
al. 1992) as `true or probable' PNe. The exception is
IRAS\,21282+5050, whose optical spectrum (Cohen \& Jones 1987) shows
[O\,{\sc iii}] emission lines, photospheric absorption features
corresponding to a heavily reddened [WC\,11] nucleus, and a total
luminosity of about 2$\times10^{3}$~L$_{\odot}$ for a guessed distance
of 2~kpc, making it a probable low-excitation PN (the central star has
been re-classified as an O star by Crowther et al. 1998).

In the notation of Table \ref{table:class}, out of 74 galactic disk
PNe with 8-13$\mu$m spectra, there are 12 O nebulae (of which SwSt1
and IC4997 also show the UIR bands), 16 c nebulae and 26 C nebulae
(for 9 of which the fits are improved with the inclusion of SiC). The
remainder shows either too little continuum emission (17 PNe), or no
clear identification in terms of the classification used here (e.g.
IC2149 and M1-14 whose spectra are best fitted by graphite emisssion
only, and K4-57, whose spectrum is flat in units of Janskys, and is
atypical of PNe). Finally the fits in Hb12 and Vy2-2, both classified
as `O' PNe, require some amount of SiC, which is probably due to
variations in the silicate emission profile rather than a
superposition of grain types.

\section{Comparison of the dust compositions and  the gas phase abundances}\label{sec:gasphase}

How does the warm dust emission feature classification, based on the
grain C/O chemical balance, compare with the gas phase C/O abundance
ratios? How does it compare to nitrogen enrichment?  In order to
address these questions, we searched the literature for published gas
phase abundances in PNe with 8-13$\mu$m spectra. Table
\ref{table:dusttypes} lists the PNe for which a detailed spectroscopic
analysis is available, but it should be borne in mind that the
uncertainties in the abundance analysis are often substantial.  The
PNe are classified according to the Peimbert (1978) types, with the
sub-types in N/O ratio introduced by Fa\'{u}ndez-Abans \& Maciel
(1987). Many assignments to N/O types are taken from the compilation
in Maciel \& Dutra (1992).

\begin{figure}
\begin{center}
\resizebox{8.5cm}{!}{\epsfig{file=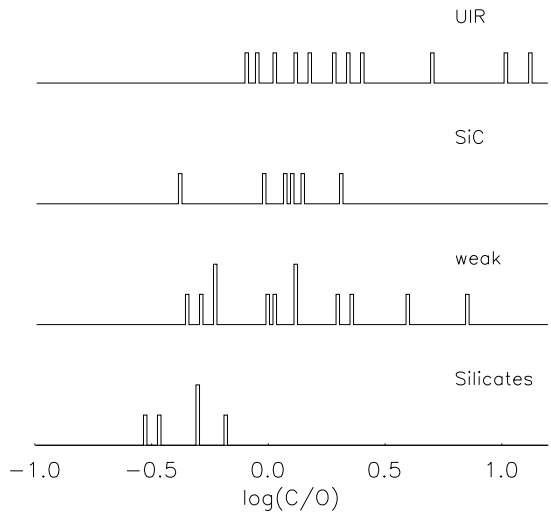}}
\end{center}
\caption{ Histogram showing the number of nebulae of each dust types
as a function of gas phase C/O ratios. Typical error bars are $\sim
+0.11/-0.15$ horizontally for a 20 per cent uncertainty in
the abundance calculations.}\label{fig:dust_CO}
\end{figure}

Figure \ref{fig:dust_CO} shows the distribution of gas phase C/O
ratios for each 8-13$\mu$m continuum class. The correspondence is not
direct, some nebulae with C based grains have gas phase C/O$<$1. There
is however a good correlation in C based grains with C/O ratio,
previously obtained by Barlow (1983) and Roche (1989), for example,
but with a lower number of PNe. Silicate grains indicate an O rich
environment, SiC a C rich environment, and the UIR bands correlate
with a strong overabundance of C relative to O.  Weak continuum PNe
are widely spread in C/O ratios, suggesting they may correspond to
later evolutionary stages, rather than C/O$\sim$1. It is established
that the dust grain composition reflects on average the gas phase
composition, and can be used as probes of C enrichment in PNe . The
warm dust features are thus an alternative to the UV lines [C\,{\sc
ii}]\,$\lambda$2326, [C\,{\sc iii}]\,$\lambda$1908, [C\,{\sc
iv}]\,$\lambda$1550, for coarse classifications of C/O ratios.

The existence of 5 PNe with C/O$<$1 and grain emission characteristic
of C rich environments merits further attention. The same apparently
paradoxical situation is hinted at by the superposition of silicates
and C based grains in the spectra of SwSt1, IC4997, Hb12 (confirmed by
measurements at 3~$\mu$m showing the 3.3~$\mu$m UIR band, Roche et
al. 1996). Standard equilibrium chemistry suggests the least abundant
of C or O would be locked in CO (Gilman 1969), and the production of
C-rich molecules in O-rich circumstellar environments may be a rather
rare non-equilibrium phenomenon (observed towards certain M
supergiants in h and $\chi$ Per, Sylvester et al. 1998). This is also
suggested by the appearance of both C- and O-type grains in novae
(e.g. Smith et a. 1995). Thus mixtures of grain types imply either a
stratification in the ejecta composition of the progenitor star, or
the mixing of the progenitor ejecta with pre-existing material. This
point is relevant when linking the grain types with the C/O ratios of
the progenitor stars, and will be discussed further in a forthcoming
article (Casassus \& Roche 2000, paper II).

\begin{figure}
\begin{center}
\resizebox{8.5cm}{!}{\epsfig{file=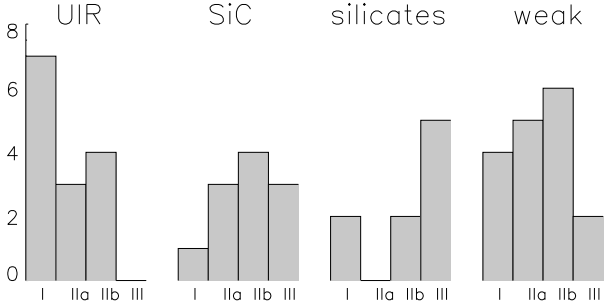}}
\end{center}
\caption{Relative proportion of PN dust types among Peimbert (1978)
types, with sub-types from Fa\'{u}ndez-Abans \& Maciel (1987).}
\label{fig:hist}
\end{figure}

The link between the dust types and N enrichment is shown in Figure
\ref{fig:hist}. The UIR bands are found mostly in nebulae of Peimbert
type I, whereas 
SiC emission and `weak' continuum PNe are uniformly
represented. Silicates are found either with strong N enrichment (type
I), or none at all (types IIb and III). These trends should be
confirmed with a larger sample of objects with known N/O ratios,
especially for O-type signatures.


The stratification of PNe in N/O ratios with height above the galactic
plane suggests that the Peimbert classification is indicative of
progenitor mass.  Thus, on a relative mass scale, the UIR bands
correspond to higher progenitor masses, SiC to intermediate masses,
silicates are found mainly for low mass progenitors, but also for the
most massive ones. The uniform distribution of `weak' continuum PNe
again suggests they may correspond to later evolutionary stages.

But the dichotomy between O- and C-type PNe for high gas phase N/O
abundance ratios indicates that there is no simple correspondence
between progenitor mass and dust signature. The dust emission features
provide complementary information to the Peimbert types.

\section{The sky distribution of PNe dust types}\label{sec:sky}

\begin{figure}
\begin{center}
\resizebox{8.5cm}{!}{\epsfig{file=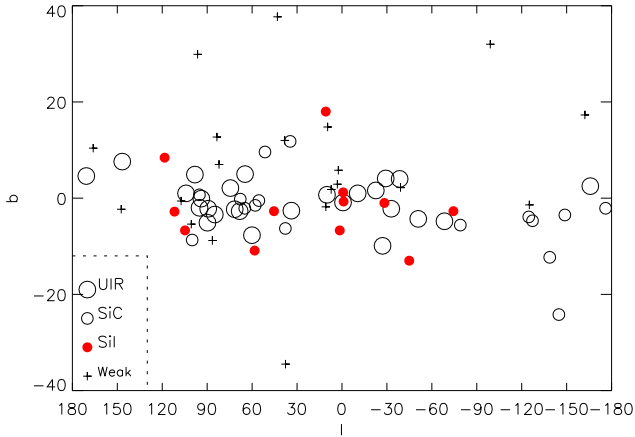}}
\end{center}
\caption{Sky distribution of PN dust types.}
\label{fig:lb.eps}
\end{figure}

The sky distribution of PN dust types, built using Table
\ref{table:dusttypes}, is shown in Figure \ref{fig:lb.eps}.  Table
\ref{table:stats_sky} contains the moments of the distribution. We
assumed the main source of errors on the fraction of each dust type is
due to `counting noise', and the uncertainties in the mean, $<b>$, and
spread, $b_\mathrm{rms}$, are estimated under the assumption that the
parent distribution of PN latitudes is normal.

\begin{table}
\begin{center}
\caption{ The properties of the  sky distribution of planetary
nebulae dust types. The errors quoted correspond to one standard deviation.}\label{table:stats_sky}
\begin{tabular}{ccccccc}  \\ \hline
 		&    $<b>$            & $b_{\mathrm{rms}}$     & N         &  $l<90$     & $90<l$                               \\
		&   [degrees]           &[degrees]             &           &   $l>270$    & $l<180$  \\ \hline
             O  &     -1.6$\pm$2.3      &   {\bf 8.0$\pm$1.6}  &  {\bf  12}&   9  &  3   \\
             c  &     -3.3$\pm$2.0      &   {\bf 7.9$\pm$1.4}  &  {\bf  16}&   8  &  8  \\
             C  &     -0.4$\pm$0.8      &   {\bf 4.1$\pm$0.6}  &  {\bf  26}&   19 &  7  \\
             +  &     6.6$\pm$3.5      &       15.5$\pm$2.5    &        20 &   12 &  8  \\   \hline
\end{tabular}
\end{center}
\end{table}

The mean $b$ is reasonably close to the galactic plane for all
identified dust types. The much larger spread in `weak' PNe suggests
they are closer on average; a larger spread in height above the plane
is discarded on the basis that `weak' PNe are uniformly distributed in
Peimbert types, although the adoption of a distance scale is required
to further discuss this issue. It seems that UIR PNe are the most
concentrated towards the galactic plane, although SiC and silicate PNe
have roughly the same spread in galactic latitude. For the subset of
PNe with C based grains (the SiC and UIR nebulae),
$b_{\mathrm{rms}}$=6.0$\pm$0.7, giving the ratio of the spreads for O
rich PNe to C rich PNe,
$b_{\mathrm{rms}}(\mathrm{O})/b_{\mathrm{rms}}(\mathrm{C}) = 1.3 \pm
0.3$. All uncertainties quoted in this article are $\pm 1\sigma$.

There is a hint of a decrease in the proportion of silicate grains PNe from
the first and fourth quadrants ($-90<l<90$) to the second and third
quadrants ($90<l<270$), from 0.25$\pm$0.07 to 0.17$\pm$0.09.  Since it
corresponds to only one $\sigma$ it cannot be considered a solid
property of the distribution. There is also an indication of an increase 
in the relative proportion of SiC PNe, which doubles from
0.22$\pm$0.07 to 0.44$\pm$0.11.

\section{Adopted distance scales}\label{sec:PNdists}

\subsection{Distance scales based on 6~cm continuum emission}

One of the persistent problems related to the study of PNe is the
difficulty of obtaining accurate distances.  A review of the methods
based on individual properties of PNe can be found in Peimbert
(1992). These so-called `direct' distance estimators are available for
a restricted number of objects, and suffer from intrinsic
uncertainties. However, the distances to PNe can be determined on a
statistical basis, which match the average properties of PNe. A
distance scale for PNe can be built on the assumption that a general
relationship holds for a set of PNe. We adopted the distance scale
derived by Zhang (1995), who used an arithmetic average of two
complementary methods, one based on the mass-radius relationship
(e.g. the review by Kwok 1994), and the other on the relationship
between $T_b$, the free-free 6~cm brightness temperature, and nebular
radius (as introduced by Van de Steene \& Zijlstra 1994). Zhang (1995)
calibrated the mass-radius and $T_b$-radius relationships with a large
sample of PNe with individually and `directly' determined
distances. This `direct' method is explained in detail in Zhang and
Kwok (1993) and Zhang (1993), and depends on the distance-independent
parameters $T_b$ and the central star temperature $T_\star$. Distances
thus obtained are strongly model dependent, and can be in disagreement
with the more accurate comparison of angular expansion rate and radial
velocity. NGC~6572, NGC~6302, NGC~3242, NGC~2392, and NGC~7662 are
given `direct' distances of, respectively, 2.9~kpc, 0.1~kpc, 1.1~kpc,
0.5~kpc, 1.6~kpc, while their expansion distance is $1.5\pm0.5$,
$1.6\pm0.6$~kpc, $0.4\pm0.1$~kpc, $>1.4$~kpc, $0.8\pm0.7$~kpc (Gomez
et al. 1993, Hajian et al. 1995, Hajian \& Terzian 1996).  But the
details of the distances to each nebula is of secondary importance as
long as the global properties of PNe are reproduced. In that sense,
the Zhang (1995) distance scale gives a Gaussian distribution about
the galactic centre for bulge PNe, with a narrower scatter than the
scale by Van de Steene \& Zijlstra (1994).

\subsection{Distances from  {\em IRAS} fluxes to optically thick PNe}

As the central stars of PNe evolve rapidly in time, and the nebulae
are active radiatively and dynamically (e.g. Kwok 1994), distance
scales based on invariant properties of PNe cannot be applied to the
whole PN population. But in the case of the sample discussed here,
the 4 $IRAS$ band fluxes, coupled with constant luminosity, may
provide an alternative distance scale, as we now argue.

The compact and IR-bright PNe are likely to be young, surrounded by
substantial molecular material and therefore optically thick to the
ionizing radiation from the central star.  In this case the total
luminosity of the nuclei can be inferred from the flux of any HI
recombination line, by equating the number of H$^{+}$ recombinations
to the number of photoionizations.  M\'{e}ndez et al. (1992) tested
this hypothesis by comparing with spectroscopic studies of PN nuclei,
linking the surface gravity and effective temperature to the
luminosity through atmosphere models. Their conclusion is that most
PNe are optically thin. However, their sample is biased against
obscured central stars: out of 23 PNe, 6 are infrared-bright and are
among the sample discussed here, 4 of which have no warm dust. Thus
M\'{e}ndez et al. (1992) included only two nebulae with 8--13$\mu$m
spectra showing warm dust emission, for which the ratio of
luminosities derived from optical thickness to the model-atmosphere
luminosities are 0.70 and 0.92 (for M1-26 and IC418). It is thus
likely that the compact and IR-bright PNe with warm dust emission are
optically thick.

In PNe which are optically thick in the Lyman continuum, the ionized
central regions are surrounded by substantial amounts of neutral gas,
an environment favourable to dust-grain survival. Most of the UV
radiation escaping from the ionized region would be absorbed by dust
grains, which heat up as a result to $\sim$100-200\,K, and re-radiate
in the mid- and far-IR spectral range. The {\em IRAS} band fluxes
should give a good representation of the bolometric fluxes using
\begin{equation}\label{eq:firas}
F_{IRAS}=\sum_{j=1}^{4} \nu I_{\nu}(j),
\end{equation}
where the sum extends to the 4 $IRAS$ bands.

\begin{figure}
\centering
\mbox{\subfigure{\epsfig{figure=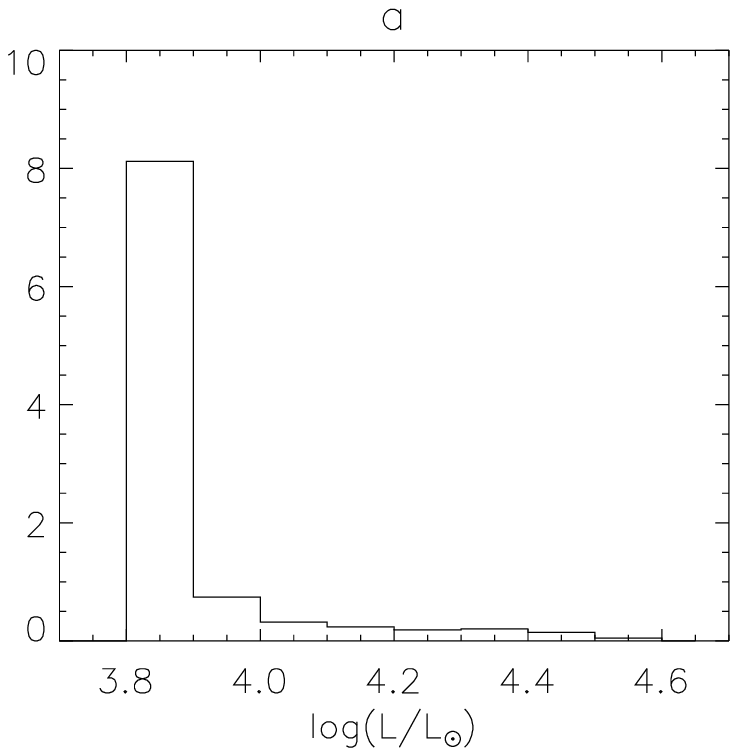,width=4.cm}}\quad
\subfigure{\epsfig{figure=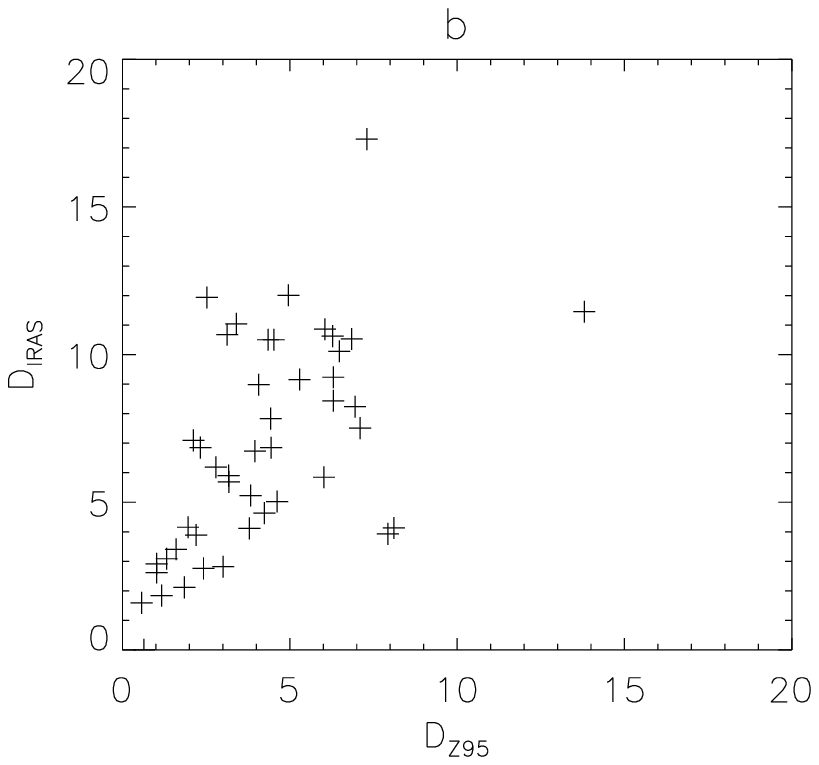,width=4.cm}}}
\caption{{\bf a)} Synthetic PN luminosity function from the
initial-final mass relationship of Groenewegen \& de Jong (1993), and
with progenitor ZAMS masses between 1.2 and 7~M$_{\odot}$. {\bf b)}
The relationship between $D_{IRAS}$ and Zhang (1995) for the PNe with
warm dust emission.}\label{fig:PNLF}
\end{figure}

PNe initially evolve at a constant luminosity once they leave the AGB,
as was first shown by Paczy\'{n}ski (1970, see also Bl\"{o}cker
1995). The luminosity function for the youngest PNe should be close to
that of tip-of-the-AGB objects.  The distribution of core masses for
stars at the tip of the AGB can be calculated with a synthetic AGB
model and a crude galactic disk model (we used the analytic
prescriptions in Groenewegen \& de Jong 1993, and a galactic disk
model described in paper~II). The core-mass luminosity relationship
from Wagenhuber \& Groenewegen (1999), in the case of post-AGB objects
(i.e. in the asymptotic regime and vanishing envelope mass), gives the
luminosity function of young PNe\footnote{PNe progenitors for the
sample discussed here were assumed to have masses in the range
1.2$<$M/M${_\odot}<$7, see paper~II.} shown in Figure \ref{fig:PNLF}a,
with a mean of 8500\,L$_{\odot}$. A very similar luminosity function,
with an average of 9300\,L$_{\odot}$, is obtained using the
prescriptions in Wagenhuber \& Groenewegen (1999) for the initial-final
mass and core-mass-luminosity relations, and taking solar metallicity
and an IMF index of 1 (instead of 1.72 in paper~II, in a notation
where the Salpeter (1955) IMF would be 1.35), with a constant star
formation rate.

It appears the PN luminosities are not expected to vary over more than
one order of magnitude.  Assigning the same luminosity of
8500~L$_{\odot}$ for all PNe gives a maximum error on the distance of
only a factor $\ls 2$. Distances to compact and IR bright PNe can thus
be estimated under the assumption of constant
luminosity. Eq. \ref{eq:firas} for the bolometric flux is likely to be
a lower limit only, but in this article we are interested in the
relative properties of the distribution of the different PNe dust
types, and their absolute distances are not required. We will refer to
distances derived in this way by $D_{IRAS}$, and those from Zhang as
$D_{Z95}$.

We stress $D_{IRAS}$ distances are only meant to investigate an
independent distance scale and its consequences on the derived
galactocentric trends; these distances should not be taken as
accurate.

\subsection{Adopted distances to compact and IR-bright PNe}

The distances derived from the two methods described above are listed
in Table~\ref{table:dusttypes}. $D_{IRAS}$ distances appear to be
reasonable for PNe with detected warm dust emission. Figure
\ref{fig:PNLF}b shows a good correlation between the two distance
estimates in the case of PNe with warm dust emission: The ratio
$D_{IRAS}/D_{Z95}$ is 1.87 on average, with a 1-$\sigma$ spread of
0.84. This suggests that the luminosity used to derive $D_{IRAS}$
distances may be overestimated by a factor $\sim3-4$, if $D_{Z95}$
distances are reliable. Also, the cases of NGC~6302, NGC~6572 and
BD+303639 (three PNe with warm dust emission features) allow comparing
their expansion distances of $1.6\pm0.6$~kpc, $1.5\pm0.5$~kpc and
$2.68\pm0.81$~kpc (Gomez et al. 1993, Hajian et al. 1995), with their
$D_{IRAS}$ distances of 1.6~kpc, 2.9~kpc and 2.12~kpc. Although the
comparison supports $D_{IRAS}$ distances, a handful of objects does
not permit a generalization. In any case, $D_{IRAS}$ may be used as an
upper limit, except for PNe with upper limits in the {\em IRAS}
100$\mu$m band. It is worth noting, however, that the PNe Vy2-2,
IRAS21282+5050, Hb12 and Hen2-113 are given distances on the Zhang
(1995) scale that are in excess of $D_{IRAS}$ by a factor larger than
1.7 (which takes into account the maximum range expected in the PN
luminosity function).


In the remainder of this article we assign $D_{IRAS}$ distances to PNe
without radio data (i.e. M2-56 and HDE330036). The case of K\,3-69
seems to be anomalous: both distance estimates give $\sim$25\,kpc,
putting K\,3-69 at 1.7\,kpc above the galactic plane, and we preferred
to use the distance of 7.9\,kpc from Cahn et al. (1992). Another
anomalous case is M2-54: again, the Zhang (1995) distance scale and
$D_{IRAS}$ both give a distance of $\sim$13\,kpc, placing it near the
northern galactic warp. In order to avoid the uncertainties associated
with exaggeratedly large distances, we adopted a maximum
galactocentric radius of 14\,kpc to compute the moments of the
vertical distribution, thus excluding K\,3-69 and M\,2-54.

\section{Tests for the completeness and homogeneity of the compact and IR-bright PN sample}\label{sec:statproof}

The sequence in Peimbert (1978) types is indicative of progenitor
mass, the highest being associated with type~I. There is a
stratification in height above the galactic plane as a function of N/O
type (e.g. Maciel and Dutra 1992). Such a stratification is indeed
present in this sample. Inside the solar circle, the root mean square
height over the plane, $z_{\mathrm{rms}}$, is 0.14\,kpc for type I,
0.37 for type IIa, 0.45 for type IIb, and 0.68 for type III. This
stratification is an indication that a statistical study based on the
compact and IR-bright PN sample would be sensitive to PN properties
with the same dependence on progenitor mass as the Peimbert types. Out
of 49 compact and IR-bright PNe with known N/O ratio, 29$\pm$6\% are
type I, 21$\pm$6\% are type IIa, which is typical of PN catalogues
(e.g. Maciel \& Dutra 1992). The disk PN population seems to be
homogeneously sampled, although the constraints will remain loose
until a larger sample is available.

A discussion of the selection effects is possible in terms of a
comparison between the fraction of O-rich PNe and the predictions of
synthetic AGB models. In paper~II we compare expected tip-of-the-AGB
statistics for the C/O chemical balance with those from the dust
signatures. We find that for a minimum PN progenitor mass of
1\,M$_{\odot}$, about 50\% of all young PNe should be O-rich, whereas
we report 22\%. To match the observed ratio, the minimum PN progenitor
mass for the sample in this work must be at least
$M^\mathrm{min}$=1.2, at a $2\sigma$ confidence level - assuming the
warm dust composition correponds to the last $\sim 2000$~yr of AGB
evolution.  Averaging the AGB ejecta over the last 25000~yr increases
the fraction of O-rich PNe by $\sim 10\%$, which may explain the
higher frequency of O-rich nebulae with plasma diagnostics ($40\pm8\%$
in the sample used here).

It is possible, however, that differences in the opacity functions
among C- or O-based grains could lead to different lifetimes of the
warm-dust emission phase. In this case the tip-of-the-AGB and observed
C/O statistics would be different. To summarise, the maximum
separation from the central star, $r_0$, required to keep a dust grain
at a temperature $T>T_0$ is $r_0 \propto \sqrt{k_\star/k_T}$, where
$k_\star$ is the opacity averaged over the central star spectrum, and
$k_T$ is averaged over a black body at $T_0$.  For instance, it may be
thought that if C-rich grains have higher $k_\star/k_T$ than O-rich
grains, then the C-rich warm-dust phase would be longer. But either A)
$k_T$ is fixed, and then the acceleration of C-rich grains by
UV-radiation pressure would be higher, thereby shortening the C-rich
warm-dust phase, or B) $k_\star$ is fixed and $k_T$ is lower for
C-rich grains, in which case the 10$\mu$m fluxes of C-rich PNe would
be lower, limiting the number of C-rich PNe in spite of their
hypothetical extended lifetimes. In addition the role of the stellar
wind and the interaction with the nebula considerably complicate the
picture.

But a test can be found for the preferential selection of O- or C-rich
grains. The good agreement between $D_{IRAS}$ and $D_{Z95}$ suggests
Eq.  \ref{eq:firas} can be used to estimate the fraction of luminosity
radiated in the 12$\mu$m {\em IRAS} band by PNe with warm dust
emission, with
\begin{equation}
\frac{L_{12{\mu}m}}{L\star}=\frac{\nu I \nu(12{\mu}m)}{\sum_{j=1}^{4} \nu I_{\nu}(j)}.
\end{equation} Omitting the PNe with upper limits in the 100$\mu$m {\em IRAS} band,
the fraction of luminosity emitted at 12$\mu$m is $\sim$0.25$\pm$0.15
for PNe with silicate emission, 0.22$\pm$0.09 for SiC PNe,
0.27$\pm$0.14 for UIR PNe, and $\sim$0.25$\pm$0.14 for all warm dust
types, while `weak' PNe have a $L_{12{\mu}m}/L\star$ ratio of
0.11$\pm$0.03 (this sample is biased towards high values of
$L_{12{\mu}m}/L\star$, so the average for all PNe would be much
lower). Considering the relatively large uncertainties, the above
values show that, for a given central star luminosity, the IR-bright
selection criterion would not preferentially select one type of grains
above others (at least in broad terms). 

At a lower flux limit of 0.5\,Jy, and under the assumption that 20\%
of the total luminosity is radiated in the 12$\mu$m {\em IRAS} band, a
good portion of the galactic disk should be sampled: the maximum
distance at which a PN with warm dust would be detected is 20\,kpc,
for L$\star$=10000~L$_{\odot}$. It is thus apparent that the
completeness and homogeneity of the sample discussed here are
dominated by the selection effects in the PN catalogues
themselves. As PNe are, for the most part, discovered through optical
surveys, the distances are unlikely to be much in excess of 3\,kpc,
especially towards the inner Galaxy.


\section{The galactic disk distribution of PN dust emission features}\label{sec:pnedistrib}

\begin{figure}
\begin{center}
\resizebox{7cm}{!}{\epsfig{file=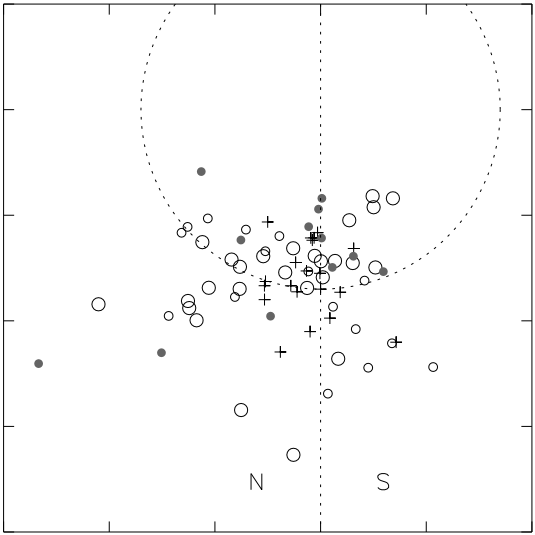}}
\caption{ The galactic disk distribution of warm-dust PNe on a face-on
view of the disk, with distances from Zhang (1995). Ticks are every
5\,kpc, and the solar circle is shown in dashed
line.}\label{fig:distr}
\end{center}
\end{figure}

The spatial distribution of PNe dust types can now be constructed
using Table \ref{table:dusttypes}, and is shown in Figure
\ref{fig:distr} in the case of the Zhang (1995) distance scale.  Table
\ref{table:stats} lists the properties of the distribution. The total
number of objects is Ntot=54, consisting of 33 for $R<R_\circ$, and 21
for $R>R_\circ$ ($R_\circ=8.5$\,kpc, Kerr \& Lynden-Bell 1986). The
decrease in the relative proportion of Silicate PNe, which was hinted
at in Section \ref{sec:sky}, is confirmed to a somewhat higher degree
of significance: the fraction of Silicate PNe decreases from
0.27$\pm$0.08 for $R<R_\circ$ to 0.14$\pm$0.08 for $R>R_\circ$. There
is a concentration of UIR PNe towards z=0, confirming that they are
related to higher progenitor masses. On the other hand SiC and
silicate nebulae have a similar spread in height above the galactic
plane. However, the vertical distribution of PN dust types is rather
homogeneous when compared to that obtained in the Peimbert types.

Table \ref{table:stats_IRAS} summarises the properties of the PN
distribution obtained with $D_{IRAS}$ distances. It may seem
surprising that SiC nebulae are at greater distances on average, but
it is compatible with finding most SiC nebulae outside the solar
circle, where PN catalogues are less affected by interstellar
extinction. The proportion of silicate PNe is confirmed to decrease
with $R$ at a higher degree of significance: the fraction of silicate
PNe decreases from $0.32\pm0.09$ for $R < R_\circ$ to 0.14$\pm$0.06
for $R>R_\circ$.


As is apparent from Figure \ref{fig:distr}, `weak' nebulae are closer
on the Zhang (1995) scale. It was mentioned that PNe with no warm dust
are equally distributed among Peimbert types, and that their gas phase
C/O ratios reflect the proportions for the whole sample. They also
have higher 6\,cm fluxes, which together with the selection criteria
of compact angular size, explains why they are on average closer. It
is thus very likely that PNe with weak continuum correspond to later
evolutionary stages, and are not a transition stage where C/O$\sim$1.
Although $D_{IRAS}$ distances are not applicable to `weak' PNe, whose
optical thickness is uncertain, it is interesting to note that the
average $D_{IRAS}$ distance to `weak' PNe is 9.0~kpc, greater than for
any other type of PNe. This could be interpreted as a lower average
luminosity (as expected for more evolved PNe on the white dwarf
cooling track), or that for `weak' PNe the far-IR flux $F_{IRAS}$ is
not a good approximation to the bolometric flux.

There is a peculiar asymmetry in the face-on distribution of PNe of
Figure \ref{fig:distr}. The sector of the galactic disk with southern
galactic longitudes (the third and fourth quadrants) is
underpopulated. This is an effect due simply to the incompleteness of
the catalogues. The same asymmetry can be seen in the face-on map of
Durand et al. (1998), with a larger number of PNe. Warm dust PNe with
reliable IRAS fluxes all gather in very tight IRAS colour-colour boxes
(in particular $\log$(F(100$\mu$m)/F(60$\mu$m))$<0$,
$\log$(F(25$\mu$m)/F(12$\mu$m))$>0$) which allows selecting all
warm-dust PN candidates from the IRAS PSC. We found 331 IRAS point
sources with colours of warm dust PNe, whose galactic
longitude/latitude distribution is uniform from northern to southern
longitudes. Also, the Carina spiral arm between the third and fourth
quadrant is viewed tangentially from the sun, thus increasing the
interstellar extinction and limiting the PN discovery rate.

\begin{table}
\begin{center}
\caption{ The properties of the distribution of PN dust types, based
on the distances from Zhang (1995). The errors quoted correspond to
one standard deviation.}\label{table:stats}
\begin{tabular}{cccccccc}  \\ \hline
 		&  {\small $D/D_\mathrm{rms}$} &  $<z>$            & $z_{\mathrm{rms}}$   &  \multicolumn{3}{l}{N}\\
		&   [kpc]               &[100\,pc]           &[100\,pc]             &           &  {\small $<R_\circ$}    & {\small $>R_\circ$}  \\ \hline
             O  &   3.9/2.1             &  -1.0$\pm$1.4      &   {\bf 4.7$\pm$1.0}  &        12 &   9  &  3   \\
             c  &   4.3/2.0             &  -1.8$\pm$1.0      &   {\bf 4.1$\pm$0.7}  &        16 &   7  &  9   \\
             C  &   3.8/2.3             &  -0.0$\pm$0.5      &   {\bf 2.3$\pm$0.3}  &        26 &  17  &  9  \\
             +  &   2.3/1.0             &   2.3$\pm$0.9      &        4.7$\pm$0.8   &        19 &  10  &  9   \\   \hline
\end{tabular}
\end{center}
\end{table}

\begin{table}
\begin{center}
\caption{ Same as table \ref{table:stats}, but with $D_{IRAS}$ distances.}\label{table:stats_IRAS}
\begin{tabular}{cccccccc}  \\ \hline
 		&  {\small $D/D_\mathrm{rms}$ } &  $<z>$            & $z_{\mathrm{rms}}$   &  \multicolumn{3}{l}{N}\\
		& [kpc]   &     [100\,pc]           &[100\,pc]             &           & {\small  $<R_\circ$}    & {\small $>R_\circ$}  \\ \hline
             O  & 4.0/1.6 &   -1.8$\pm$1.9      &   {\bf 6.2$\pm$1.3}  &        12 &   8  &  4   \\
             c  & 8.4/3.6 &   -4.6$\pm$2.7      &   {\bf 11.0$\pm$1.9} &        16 &   4  &  12  \\
             C & 5.6/3.2 &   -0.7$\pm$0.9      &   {\bf 4.4$\pm$0.6}  &        26 &  13  &  13  \\  \hline
\end{tabular}
\end{center}
\end{table}

\section{Conclusions}\label{sec:conclusion_chap3}

The total number of PNe with 8--13$\mu$m spectra has been increased to
74 with the inclusion of 24 new objects. The sample consists of
compact and IR-bright galactic disk PNe listed in the Strasbourg-ESO
catalogue. 54 PNe have clearly identified warm dust emission features,
which are placed into three groups (see Table~\ref{table:class}): 12
PNe show silicate emission, 16 show SiC, and 26 show the UIR
bands. The remainder have 8--13$\mu$m spectra dominated by emission
lines, and correspond to later evolutionary stages.  Thus 22$\pm$6\%
of the PNe with warm dust emission have O-rich grains. We have used
this sample for an initial study of the PN dust emission features in
the galactic context.

A comparison of the PNe dust types with the gas phase C/O ratio shows
a good correspondence: Silicate nebulae have C/O$<$1, SiC nebulae are
found with C/O$\gs$1, while PNe that show the UIR bands often have
C/O$>>$1. We thus confirm that the dust emission features 
represent an alternative to the plasma diagnostic for measuring the
C/O chemical balance in PNe. Nebulae that show the UIR emission bands
also have the highest N/O gas phase ratio. Silicate nebulae are found
either with high N/O ratios, or no nitrogen enrichment at all. On the
other hand SiC nebulae are more uniformly distributed in N/O ratios.
Thus, on a relative mass scale, PNe with emission from the UIR bands
correspond to higher progenitor masses, and those with SiC to intermediate
masses. Silicates are found
mainly for low mass progenitors, but also for the most massive
ones. The dust emission features thus provide complementary
information on the progenitors masses to the Peimbert types.

The adoption of statistical distances showed that the sample is large
enough to show stratification in Peimbert types. We find a link
between objects with UIR band emission and higher progenitor masses,
as indicated by their concentration towards the galactic plane,
obtained from their sky distribution and through the use of two
independent PN distance scales. There is a trend for a decreasing
proportion of O-rich PNe with galactocentric radius, confirmed by both
distance scales, from 30$\pm$9\% inside the solar circle, to
14$\pm$7\% outside.  This trend reflects the variations in the M/C
star ratio from Thronson et al. (1987) and Jura et al. (1989).



We also showed that the {\em IRAS} fluxes are a good representation of
the bolometric flux for PNe with warm-dust emission
(Section~\ref{sec:PNdists}). The requirement $F(12\mu$m$)\,>\,0.5\,$Jy
should probe a good portion of the galactic disk, and the dominant
selection effects are rooted in the PNe catalogues.

Although most known IR bright and compact PNe were included in this
study, further observations are required to improve the
statistics. Large aperture telescopes and mid-IR array detectors will
be much more sensitive for the detection of the warm dust emission
features in these compact objects, and could allow a more accurate
analysis.


\section*{Acknowledgments}
We are grateful to the referee for interesting comments, and to PATT
for the time allocation on UKIRT, which is operated by the Joint
Astronomy Centre on behalf of the Particle Physics and Astronomy
Research Council, and PATT and ATAC for allocations on the AAT.
S.C. acknowledges support from Fundaci\'{o}n Andes and PPARC through a
Gemini studentship.

\bsp
\label{lastpage}
\end{document}


\small
\setcounter{table}{3}
\begin{center}
\begin{sidewaystable}
\caption{List of all PNe with 8-13$\mu$m spectra, and the identified warm dust emission features.}\label{table:dusttypes}
\begin{tabular}{lcccclcccccccl}   \\ \hline
	name	&  $l$	 & $b$   & dust	   &ref.    &N/O   &He     &C         &N        &O        &ref. &$D$ & $D$       &    comments	                        \\
               	&      	 &     	 & type	   &$^a$    &type  &       &          &         &         &$^h$ &Z95 & {\em IRAS}&      	    	       	        \\
      solar     &        &       &         &        &      &  0.10 &  3.6(-4) & 9.3(-5) & 7.4(-4) & 24  &    &           &                                     \\
          SwSt1	&  1.50	 & -6.70 &    O    &   3    &  III &  0.11 &  3.9(-4) & 4.1(-5) & 5.9(-4) & 1   &3.79& 4.12 &silicates:0.95,203  UIR:0.05,2000    \\
        NGC6369	&  2.40  &  5.80 &    +    &   4    &  I   &  0.13 &  5.8(-4) & 1.2(-4) & 2.9(-4) & 2   &0.75& 3.96 &weak continuum                       \\
            Hb4	&  3.10  &  2.90 &    +    &   7    &  I   &  0.13 &  --      & 2.8(-4) & 4.8(-4) & 2   &2.68& 9.52 &weak continuum	                  \\
            Hb6 &  7.20  &  1.80 &    +    &   7    &  I   &  0.11 &  --      & 4.5(-4) & 5.1(-4) & 2   &2.45& 6.99 &weak continuum	    \\
        NGC6309 &  9.60  & 14.80 &    +    &   4    &  II  &  0.11 &  1.3(-3) & 1.2(-4) & 9.5(-4) & 3   &2.35& 8.16 &weak continuum	    \\
   NGC6537$^{i}$&  10.01 & 00.7  &    C    &   --   &  I   &  0.19 &  4.0(-5) & 8.9(-4) & 1.7(-4) & 23  &1.60& 3.41 &graphite:0.20,157 UIR:0.56,0.12,0.12$^{f}$  \\
        NGC6578 &  10.80 & -1.80 &    +    &   --   &  II  &  0.11 &  --      & 1.1(-4) & 5.6(-4) & 2   &2.47& 3.93 &weak continuum$^e$	    \\
           M2-9 &  10.80 & 18.00 &    O    &   3    &  III &  0.08 &  --      & 5.0(-5) & 2.3(-4) & 4   &3.01& 2.82 &thick-silicates:0.38,536,1.09 graphite:0.62,235\\
        NGC6741 &  33.80 & -2.60 &    C    &   7    &  I   &  0.11 &  8.0(-4) & 2.4(-4) & 5.4(-4) & 5   &2.33& 6.85 &graphite:0.29,128 UIR:0.71,295   \\
        NGC6572 &  34.60 & 11.80 &    c    &   3    &  II  &  0.09 &  1.3(-4) & 9.6(-5) & 3.1(-4) & 6a  &1.02& 2.91 &SiC:0.18,190 graphite:0.82,183 	    \\
        NGC7009 &  37.70 &-34.50 &    +    &   4    &  II  &  0.10 &  2.7(-4) & 1.8(-4) & 5.3(-4) & 7   &1.09& 4.35 &weak continuum	    \\
        NGC6790 &  37.80 & -6.30 &    c    &   1    &  IIb &  0.11 &  7.2(-4) & 5.5(-5) & 7.6(-4) & 8   &3.18& 5.69 &SiC:0.31,142 graphite:0.69,188	    \\
          CN3-1 &  38.20 & 12.00 &    +    &   7    &  III &  --   &  5.0(-4) & 4.9(-5) & 1.3(-4) & 3   &4.05&10.27 &weak continuum	    \\
        NGC6210 &  43.10 & 37.70 &    +    &   4    &  IIb &  0.10 &  --      & 4.9(-5) & 5.0(-4) & 3   &1.73& 6.84 &weak continuum	    \\
          Vy2-2 &  45.40 & -2.70 &    O    &   3    &  I   &  0.09 &  --      & 4.4(-5) & 2.7(-4) & 4   &7.93& 3.93 &silicates:0.95,365 SiC:0.03,100 graphite:0.02,111\\
          Hu2-1 &  51.40 &  9.60 &    c    &   7    &  III &  0.09 &  4.2(-4) & 4.0(-5) & 3.0(-4) & 3   &4.52&10.50 &SiC:0.17,200 graphite:0.83,175 \\
          M1-71 &  55.50 & -0.50 &    c    &   --   &  --  &  --   &  --      &  --     &  --     & --  &3.17& 5.90 &graphite:0.73,165 SiC:0.27,186	    \\
       Hen2-447 &  57.90 & -1.50 &    c    &   --   &  --  &  --   &  --      &  --     &  --     & --  &6.30& 8.44 &graphite:0.73,186 SiC:0.27,179          \\
         IC4997 &  58.30 &-10.90 &    O    &   3    &  III &  0.13 &  4.1(-4) & 2.0(-4) & 1.2(-3) & 6b  &4.43& 7.83 &silicates:0.95,217 UIR:0.05,2000\\
        NGC6886 &  60.10 & -7.70 &    C    &   7    &  II  &  0.09 &  4.1(-4) & 1.3(-4) & 3.1(-4) & 9   &3.13&10.67 &graphite:0.22,132 UIR:0.78,355	    \\
      BD+303639 &  64.70 &  5.00 &    C    &   3    &  IIb &  --   &  4.0(-4) & 1.1(-4) & 3.7(-4) & 10  &1.85& 2.12 &SiC:0.03,80 graphite:0.60,252  UIR:0.36,633	    \\
          K3-53 &  64.90 & -2.10 &    c    &   --   &  --  &  --   &  --      & --      & --      & --  &6.95& 8.24 &graphite:0.70,178 SiC:0.30,316	    \\
          K3-52 &  67.90 & -0.20 &    c    &   --   &  --  &  --   &  --      & --      & --      & --  &7.10& 7.51 &graphite:0.78,207 SiC:0.22,262	    \\
       Hen2-459 &  68.30 & -2.70 &    C    &   7    &  --  &  --   &  --      & --      & --      & --  &6.02& 5.85 &graphite:0.60,150 UIR:0.40,550    \\
          M3-35 &  71.60 & -2.30 &    C    &   --   &  --  &  --   &  --      & --      & --      & --  &4.44& 6.85 &graphite:0.24,191 UIR$^d$:0.57,0.07,0.12	    \\
        NGC6881 &  74.50 &  2.10 &    C    &   7    &  I   &  0.11 &  --      & 3.8(-4) & 5.5(-4) & 2   &3.96& 6.73 &graphite:0.70,150 UIR:0.30,250	    \\
        NGC6884 &  82.10 &  7.00 &    +    &   7    &  IIb &  0.10 &  1.0(-3) & 3.0(-4) & 1.0(-3) & 11  &2.63& 9.23 &weak continuum	    \\
        NGC6826 &  83.50 & 12.70 &    +    &   4    &  IIb &  0.11 &  1.7(-3) & 3.0(-5) & 2.4(-4) & 3   &1.42& 5.52 &weak continuum	    \\
        NGC7027 &  84.90 & -3.40 &    C    &   1    &  II  &  0.11 &  6.9(-4) & 1.3(-4) & 3.1(-4) & 12  &0.64&  -- &SiC:0.08,151 graphite:0.45,168 UIR:0.47,228 \\
          Hu1-2 &  86.50 & -8.80 &    +    &   --   &  I   &  0.16 &  1.2(-4) & 3.2(-4) & 1.1(-4) & 3   &2.66&17.49 &weak continuum	    \\
          M1-77 &  89.30 & -2.20 &    C    &   --   &  --  &  --   &   --     & --      & --      & --  &5.29& 9.16 &graphite:0.07,200 SiC:0.006,81 UIR$^d$:0.64,0.22,0.06\\   \hline
\end{tabular}
\end{sidewaystable}
\end{center}
\setcounter{table}{3}
\begin{flushleft}
\begin{sidewaystable}
\caption{\em continued}
\begin{tabular}{lcccclcccccccl}   \\ \hline
	name	&  $l$	 & $b$   & dust	   &ref.    &N/O   &He     &C         &N        &O        &ref. &   $D$  & $D$       &    comments	                        \\
               	&      	 &     	 & type	   &$^a$    &type  &       &          &         &         &$^h$ &   Z95  & {\em IRAS}&      	    	       	        \\
         IC5117 &  89.80 & -5.10 &    C    &   3    &  II  &  0.11 &  7.9(-4) & 1.1(-4) & 4.1(-4) & 3   &  3.83  &   5.23 &SiC:0.09,159 graphite:0.61,213 UIR:0.30,369\\
     21282$^{g}$&  93.90 & -0.10 &    C    &   8    &  --  &  --   &  --      & --      & --      & --  & 10.53  &   3.38 &prominent UIR bands\\
          M2-49 &  95.10 & -2.00 &    C    &   --   &  --  &  --   &  --      & --      & --      & --  &  6.30  &   9.23 &graphite:0.12,330 SiC:0.01,70  UIR$^d$:0.34,0.2,0.31	\\
          K3-62 &  95.20 &  0.70 &    c    &   --   &  --  &  --   &  --      & --      & --      & --  &  4.07  &   8.98 &graphite:0.79,224 SiC:0.21,89  \\
        NGC6543 &  96.40 & 29.90 &    +    &   4    &  IIb &  0.11 &  1.3(-3) & 8.7(-5) & 5.6(-4) & 3   &  1.12  &   3.36 &weak continuum	    \\
          K3-60 &  98.20 &  4.90 &    C    &   --   &  I   &  0.13 &  --      & 1.9(-4) & 4.3(-4) & 2   &  6.28  &  10.63 &graphite:0.06,106 UIR$^d$:0.59,0.17,0.18	    \\
          Me2-2 &  100.00& -8.70 &    c    &   --   &  I   &  0.14 &  --      & 6.9(-4) & 2.1(-4) & 2   &  7.30  &  17.30 &graphite:0.85,192 SiC:0.15,142	    \\
         IC5217 &  100.60& -5.40 &    +    &   7    &  --  &  --   &  --      & --      & --      & --  &   --   &   --   &weak continuum     \\
          Bl2-1 &  104.10&  1.00 &    C    &   --   &  --  &  --   &  --      & --      & --      & --  &  6.05  &  10.86 &graphite:0.06,135 UIR$^d$:0.64,0.18,0.12	    \\
          M2-54 &  104.80& -6.70 &    O    &   --   &  --  &  --   &  --      & --      & --      & --  & 13.80  &  11.45 &silicates:1.0,173	    \\
          K4-57 &  107.40& -0.60 &    +    &   --   &  --  &  --   &  --      & --      & --      & --  &   --   &   --   &unidentified	    \\
           Hb12 &  111.80& -2.80 &    O    &   3    &  IIb &  0.10 &  1.1(-4) & 6.0(-5) & 2.2(-4) & 13  &  8.11  &   4.13 &silicates:0.71,304 SiC:0.23,2000 graphite:0.05,2000\\
          M2-56 &  118.40&  8.42 &    O    &   3    &  --  &  --   &  --      & --      & --      & --  &   --   &   2.69 &thick-silicates:0.53,279,1.32 graphite:0.47,2000\\
          M4-18 &  146.70&  7.60 &    C    &   3    &  --  &  --   &  --      & --      & --      & --  &  6.85  &  10.53 &SiC:0.03,232 graphite:0.43,137 UIR:0.54,113	    \\
           M1-4 &  147.40& -2.30 &    +    &   --   &  IIb &  0.10 &  --      & 9.1(-5) & 3.2(-4) & 2   &  3.53  &  14.84 &weak continuum	    \\
         IC2149 &  166.10& 10.40 &    +    &   --   &  II  &  0.09 &  1.3(-4) & 4.3(-5) & 2.9(-4) & 14  &  2.07  &   8.85 &unidentified (graphite at 157K)   \\
          K3-69 &  170.70&  4.60 &    C    &   --   &  --  &  --   &  --      & --      & --      & --  & 25.50  &  23.23 &graphite:0.11,200 SiC:0.02,108  UIR$^d$:0.48,0.21,0.18\\
           M1-5 &  184.00& -2.10 &    c    &   --   &  --  &  --   &  --      & --      & --      & --  &  4.96  &  12.01 &graphite:0.65,187 SiC:0.35,166  \\
           J900 &  194.20&  2.50 &    C    &   3    &  IIb &  0.12 &  2.0(-3) & 4.3(-5) & 3.9(-4) & 15  &  3.40  &  11.04 &graphite:0.34,143 UIR:0.66,133	    \\
        NGC2392 &  197.80& 17.30 &    +    &   4    &  II  &  --   &  2.1(-4) & 2.1(-4) & 3.6(-4) & 16  &  1.44  &   9.84 &weak continuum	    \\
           M1-6 &  211.20& -3.50 &    c    &   7    &  IIb &  0.06 &  --      & 3.9(-5) & 3.4(-4) & 4   &  4.35  &  10.50 &SiC:0.21,110 graphite:0.79,175	    \\
          IC418 &  215.20&-24.20 &    c    &   1    &  IIb &  0.07 &  2.7(-4) & 8.3(-5) & 2.1(-4) & 6c  &  1.02  &   2.62 &SiC:0.26,120 graphite:0.74,200\\
         IC2165 &  221.30&-12.30 &    c    &   5    &  II  &  0.10 &  4.1(-4) & 7.9(-5) & 2.0(-4) & 17  &  2.52  &  11.94 &SiC$^c$:-,- graphite:-,90 	    \\
          M1-11 &  232.80& -4.70 &    c    &   3    &  III &  0.02 &  --      & 2.7(-5) & 5.5(-5) & 18  &  4.24  &   4.63 &SiC:0.25,234 graphite:0.59,194 \\
          M1-14 &  234.90& -1.40 &    +    &   --   &  III &  0.10 &  --      & 2.6(-5) & 1.9(-4) & 19  &  4.37  &  13.57 &weak continuum (possible SiC)  \\
          M1-12 &  235.30& -3.90 &    c    &   7    &  III &  0.02 &  --      & 3.2(-5) & 1.3(-4) & 18  &  6.48  &  10.11 &SiC:0.24,140 graphite:0.76,210 \\
        NGC3242 &  261.00& 32.00 &    +    &   4    &  IIb &  0.09 &  2.7(-4) & 8.1(-5) & 4.6(-4) & 3   &  0.94  &   5.51 &weak continuum	    \\
         IC2501 &  281.00& -5.60 &    c    &   3    &  II  &  0.11 &  6.1(-4) & 1.6(-4) & 5.1(-4) & 1   &  2.12  &   7.10 &SiC:0.17,137 graphite:0.83,202\\
        Hen2-47 &  285.60& -2.70 &    O    &   3    &  IIb &  0.03 &  --      & 2.3(-4) & 1.2(-3) & 4   &  3.08  &   5.66 &silicates:0.66,242 graphite:0.34,143	\\   \hline
\end{tabular}
\end{sidewaystable}
\end{flushleft}
\setcounter{table}{3}
\begin{flushleft}
\begin{sidewaystable}
\caption{\em continued}
\begin{tabular}{lcccclcccccccl}   \\ \hline
	name	&  $l$	 & $b$   & dust	   &ref.    &N/O   &He     &C         &N        &O        &ref. &   $D$  & $D$       &    comments	                        \\
               	&      	 &     	 & type	   &$^a$    &type  &       &          &         &         &$^h$ &  Z95  & {\em IRAS}&      	    	       	        \\
         IC2621 &  291.60& -4.80 &    C    &   3    & --   &  0.09 &  --      &  --     & 8.3(-4) & 15  &  2.79  &       6.19 &SiC:0.1,144 graphite:0.5,256 UIR:0.4,211 \\
        NGC5315 &  309.10& -4.30 &    C    &   3    &  I   &  0.09 &  1.6(-3) & 6.1(-4) & 6.4(-4) & 15  &  1.96  &       4.15 &SiC:0.16,2000 graphite:0.55,271 UIR:0.29,125\\
       Hen2-131 &  315.10&-13.00 &    O    &   3    &  III &  --   &  1.4(-4) & 1.8(-4) & 4.7(-4) & 1   &  2.20  &       3.89 &silicates:0.42,409 graphite:0.58,162\\
       Hen2-117 &  320.90&  2.20 &    +    &   --   & --   &  --   &  --      &  --     &  --     & --  &  --    &       --   &weak continuum (possible UIR)   \\
       Hen2-113 &  321.50&  4.00 &    C    &   2    &  IIb &  --   &  5.0(-3) & 6.6(-5) & 4.8(-4) & 22  &  5.50$^{j}$&   1.95 &UIR bands (like NGC\,7027)	    \\
       Hen2-142 &  327.10& -2.20 &    C    &   --   & --   &  --   &  --      &  --     & --      & --  &  4.62  &       5.02 &graphite:0.11,140 UIR$^d$:0.64,0.16,0.09 \\
      HDE330036 &  330.80&  4.10 &    C    &   6    &  I   &  --   &  1.9(-4) & 2.4(-4) & 2.3(-4) & 20  &  --    &       5.05 &UIR bands \\
            Mz3 &  331.70& -1.00 &    O    &   3    &  I   &  0.18 &  --      & 3.5(-4) & 2.8(-4) & 16  &  1.17  &       1.84 &thick-silicates:0.83,525,1.42 graphite:0.17,220\\
     CPD-568032 &  332.92& -9.91 &    C    &   2    &  IIb &  --   &  6.3(-3) & 8.3(-5) & 4.8(-4) & 22  &  --    &       1.85 &UIR bands (like NGC\,7027)	    \\
          Pe1-7 &  337.40&  1.60 &    C    &   --   & --   &  --   &  --      &  --     & --      & --  &  3.53$^{j}$&   3.89 &graphite:0.18,152 UIR$^d$:0.60,0.11,0.11	    \\
   NGC6302$^{i}$&  349.50&  1.00 &    C    &   7    &  I   &  0.18 &  1.0(-4) & 8.3(-4) & 5.0(-4) & 21  &  0.57  &       1.59 &graphite:0.66,154 UIR:0.34,490	    \\
          M1-26 &  358.90& -0.70 &    O    &   1    &  III &  0.08 &  1.5(-4) & 4.5(-5) & 3.0(-4) & 15  &  2.42  &       2.76 &Silicates:1,201	    \\
          19w32 &  359.20&  1.20 &    O    &   7    & --   &  --   &  --      &  --     & --      & --  &  4.30$^{j}$&   4.38 &blackbody$^b$:1,390,1.26 	    \\
            Hb5 &  359.30& -0.90 &    C    &   3    &  I   &  0.14 &  --      & 1.1(-3) & 6.6(-4) & 2   &  1.32  &       3.09 &graphite:0.41,162 UIR:0.59,248\\   \hline
\end{tabular}

\medskip

$^a$ the references for the 10$\mu$m spectra are: 1) Aitken et
al. 1979; 2) Aitken et al. 1980; 3) Aitken \& Roche 1982; 4) Roche et
al. 1983b; 5) Roche \& Aitken 1983; 6) Roche et al. 1983a; 7) Roche \&
Aitken 1986; 8) Roche et al. 1991.

$^b$ Black body with optically thick foreground silicates with
$\tau$(10$\mu$m)=1.26

$^c$ free-free continuum subtracted, SiC and graphite are required
but in a range of proportions and temperatures.

$^d$ The UIR bands were fitted using individual resonance profiles.

$^e$ The spectrum of NGC~6578 presented here, with little continuum,
is in contrast with that obtained with the {\em IRAS} LRS (Kwok et
al. 1986), which shows prominent silicate emission. Observations at
UKIRT show that a star $\sim$30~$''$ to the SE of NGC~6578 has a
silicate emission band which has probably contaminated the {\em IRAS}
LRS spectrum.

$^f$ This CGS3 spectrum was obtained with a 4$''$ beam, which may
account for the relative weakness of the 11.3$\mu$m feature compared
to other nebulae (observed with 10$''$ beams).


$^{g}$ IRAS~21282+5050. The radio data for the $S(6\,$cm)-based
distance was taken from Likkel et al. (1994).

$^{h}$ The references for the gas phase abundances are: 1) Perinotto
(1991); 2) Aller \& Keyes (1987); 3) Aller \& Czyzak (1983); 4)
K\"{o}ppen et al. (1991); 5) Hyung \& Aller (1997); 6a) Hyung et
al. (1994a); 6b) Hyung et al. (1994b); 6c) Hyung et al. (1994c); 7)
Hyung \& Aller (1995); 8) Aller et al. (1996); 9) Hyung et al. (1995);
10) Aller \& Hyung (1995);11) Hyung et al. (1997); 12) Keyes et
al. (1990); 13) Hyung \& Aller (1996); 14) Feibelman et al. (1994);
15) Kingsburgh \& Barlow (1994); 16) Zuckerman \& Aller (1986); 17)
Hyung (1994); 18) Cuisinier et al. (1996); 19) Costa et al. (1996);
20) Lutz (1984); 21) Aller et al. (1981); 22) De Marco et al. (1997).;
23) Feibelman et al. (1985); 24) Grevesse \& Anders (1989)

$^{i}$ The C abundances listed for NGC\,6302 and NGC\,6537 are
probably underestimated: in NGC~6302, C\,{\sc v} is the most abundant
ion, but was not included by Aller et al. (1981). We adopted C/O=0.9
for NGC~6302 (Mike Barlow, private communication), and did not use the
C/O information for NGC~6537.

$^{j}$ Radio data from the Strasbourg-ESO catalogue and references therein.

\end{sidewaystable}
\end{flushleft}